\documentclass[twocolumn]{webofc}
\usepackage[varg]{txfonts} 
\usepackage{hyperref}
\usepackage{amsmath}
\usepackage{amssymb}
\usepackage{bm}
\usepackage{graphicx}
\usepackage{url}
\usepackage[usenames,dvipsnames]{xcolor}
\usepackage{aas_macros}
\usepackage{rviewport}   
\hypersetup{
    bookmarks=true,         
    bookmarksopen=true,     
    bookmarksopenlevel=1,
    unicode=false,          
    pdftoolbar=true,        
    pdfmenubar=true,        
    colorlinks=true,        
    linkcolor=darkBlue,     
    citecolor=darkBlue,        
    filecolor=darkBlue,      
    urlcolor=darkBlue           
}
 \newcommand{\HI}{H\textsc{i}}
\newcommand\sfrac[2]{{\textstyle{\frac{#1}{#2}}}}
\newcommand\BB{{\overline{B}}}

\newcommand\bb{{b}}

\newcommand{\RM}{{\rm RM}}

\DeclareMathAlphabet\mathbfcal{OMS}{cmsy}{b}{n}

\def\EeV{\ifmmode {\text{Ee\kern -0.07em V}}\else
                   \text{Ee\kern -0.07em V}\fi\xspace}
\def\PeV{\ifmmode {\text{Pe\kern -0.07em V}}\else
                   \text{Pe\kern -0.07em V}\fi\xspace}
\def\TeV{\ifmmode {\text{Te\kern -0.07em V}}\else
                   \text{Te\kern -0.07em V}\fi\xspace}
\def\GeV{\ifmmode {\text{Ge\kern -0.07em V}}\else
                   \text{Ge\kern -0.07em V}\fi\xspace}
\def\eV{\ifmmode {\text{\ e\kern -0.07em V}}\else
  \text{e\kern -0.07em V}\fi\xspace}
\newcommand{\energyDensity}[1]{\ensuremath{#1\,\!\eV/\text{cm}^3}\xspace}

\definecolor{rossoCP3}{cmyk}{0,.88,.77,.40}
\definecolor{darkBlue}{rgb}{0, 0, 0.8}
\begin{document}
\title{Progress in the Global Modeling of the Galactic Magnetic Field}
\author{\firstname{Michael} \lastname{Unger}\inst{1}\fnsep\thanks{\href{mailto:Michael.Unger@kit.edu}{Michael.Unger@kit.edu}} \and
        \firstname{Glennys} \lastname{Farrar}\inst{2}\fnsep\thanks{\href{mailto:Glennys.Farrar@nyu.edu}{Glennys.Farrar@nyu.edu}}
}

\institute{Institute for Nuclear Physics, Karlsruhe Institute of Technology, Karlsruhe, Germany
\and
          Center for Cosmology and Particle Physics, New York University, New York, USA  
          }

\abstract{We discuss the global modeling of the properties of the
  Galactic Magnetic Field (GMF). Several improvements and variations
  of the model of the GMF from Jansson \& Farrar (2012) (JF12) are
  investigated in an analysis constrained by all-sky rotation measures
  of extragalactic sources and polarized and unpolarized synchrotron
  emission data from WMAP and Planck. We present the impact of the
  investigated model variations on the propagation of ultrahigh-energy
  cosmic rays in the Galaxy.} \maketitle
\section{Introduction}
\label{intro}
Magnetic fields are a major constituent of the interstellar medium of
galaxies. Their energy density in the Galactic plane is about
\energyDensity{0.5} and thus comparable to the energy density of
cosmic rays (0.8-\energyDensity{1.0}~\cite{2016ApJ...831...18C}) and
interstellar radiation fields (star light and CMB, each contributing
about \energyDensity{0.25}). The Galactic magnetic field (GMF) plays
therefore an important role for the transport and energy loss of
Galactic cosmic rays. Furthermore, it is a major nuisance for the
study of the origin of ultrahigh-energy cosmic rays as it deflects the
arrival directions of extragalactic charged particles on their way
through the Galaxy to Earth. The origin of the large-scale GMF is
still not fully understood, but it is 
widely assumed that it is
produced by the transfer of mechanical into magnetic energy via
the dynamo mechanism (see e.g.~\cite{2004astro.ph.11739S}).

Different observational tracers of the GMF are summarized
in~\cite{neronovUHECR}. In the following we focus on\\
(a) Multi-frequency observations of the Faraday
    rotation of extragalactic radio sources. The corresponding
    {\itshape rotation measures} (RMs) are proportional to the
    integral of the magnetic field parallel to the line of sight
    weighted with the density of thermal electrons of the warm ionized
    medium of the Galaxy.\\
(b) Measurements of the polarized synchrotron emission of
    cosmic-ray electrons. 
    The total polarized intensity (PI) is proportional to the
    integral of the ordered component of the
    magnetic field perpendicular to the line of sight and
    weighted by the density of cosmic-ray
    electrons. The direction of the transverse magnetic field
    component can be inferred from the Stokes parameters Q and U.\\
(c) Measurements of the total (polarized and unpolarized)
    synchrotron intensity I, which gives a measure of the line-of-sight integral
    of the product of cosmic-ray electron density and total
    magnetic field strength (perpendicular coherent field and unordered,
    random magnetic field).

The most complete attempt to determine the global structure of the
Galactic magnetic field (GMF) from these observations is the model of
Jansson \& Farrar~\cite{2012ApJ...757...14J, 2012ApJ...761L..11J}
(JF12). In this model, the GMF is described by a superposition of
three divergence-free large-scale regular components: a spiral disk
field, a toroidal halo field and a poloidal field (``X-field''). In
addition, there is a turbulent field model whose disk component is
following the same spiral structure as the regular component and whose
extended halo field strength is modeled as a Gaussian in height the
Galactic disk and as an exponential in Galactic radius.  Account is taken of the possibility of the turbulent field having a preferred direction through a striation parameter.
The parameters
of the model are derived by minimizing the $\chi^2$ between synthetic
sky maps of RM, Q, U and I and the observations.

In the following we will describe our ongoing work to improve the JF12
model and to study uncertainties in determination of its
parameters. Further 
information about the model fitting and calculation of
simulated sky maps can be found in~\cite{Unger:2017wyw}.

\section{Variations and Improvements of the JF12 Model of the Large-Scale GMF}
\subsection{Smooth Spiral Field}
\begin{figure}[!t]
    \centering
    \includegraphics[width=0.49\linewidth]{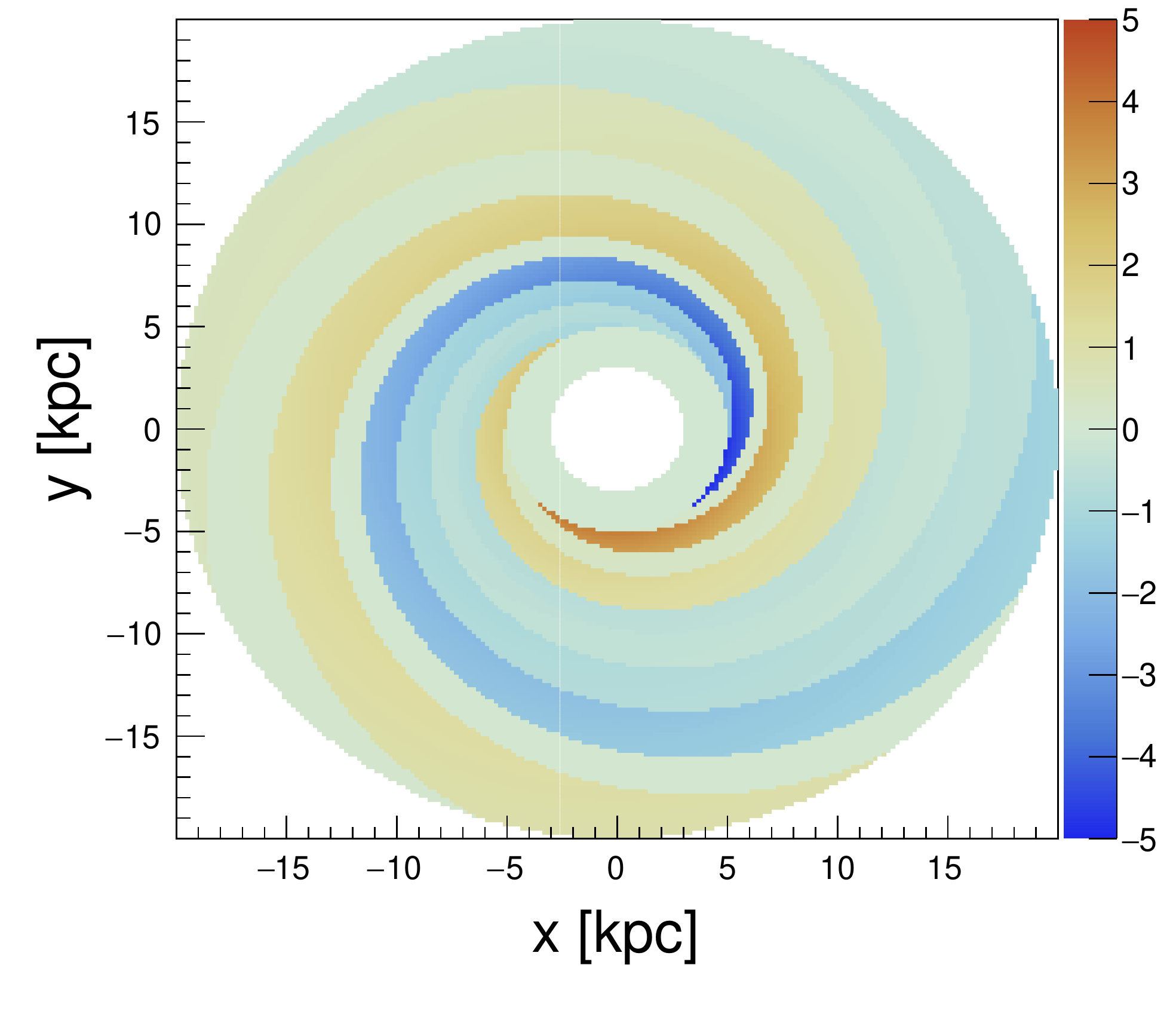}
    \includegraphics[width=0.49\linewidth]{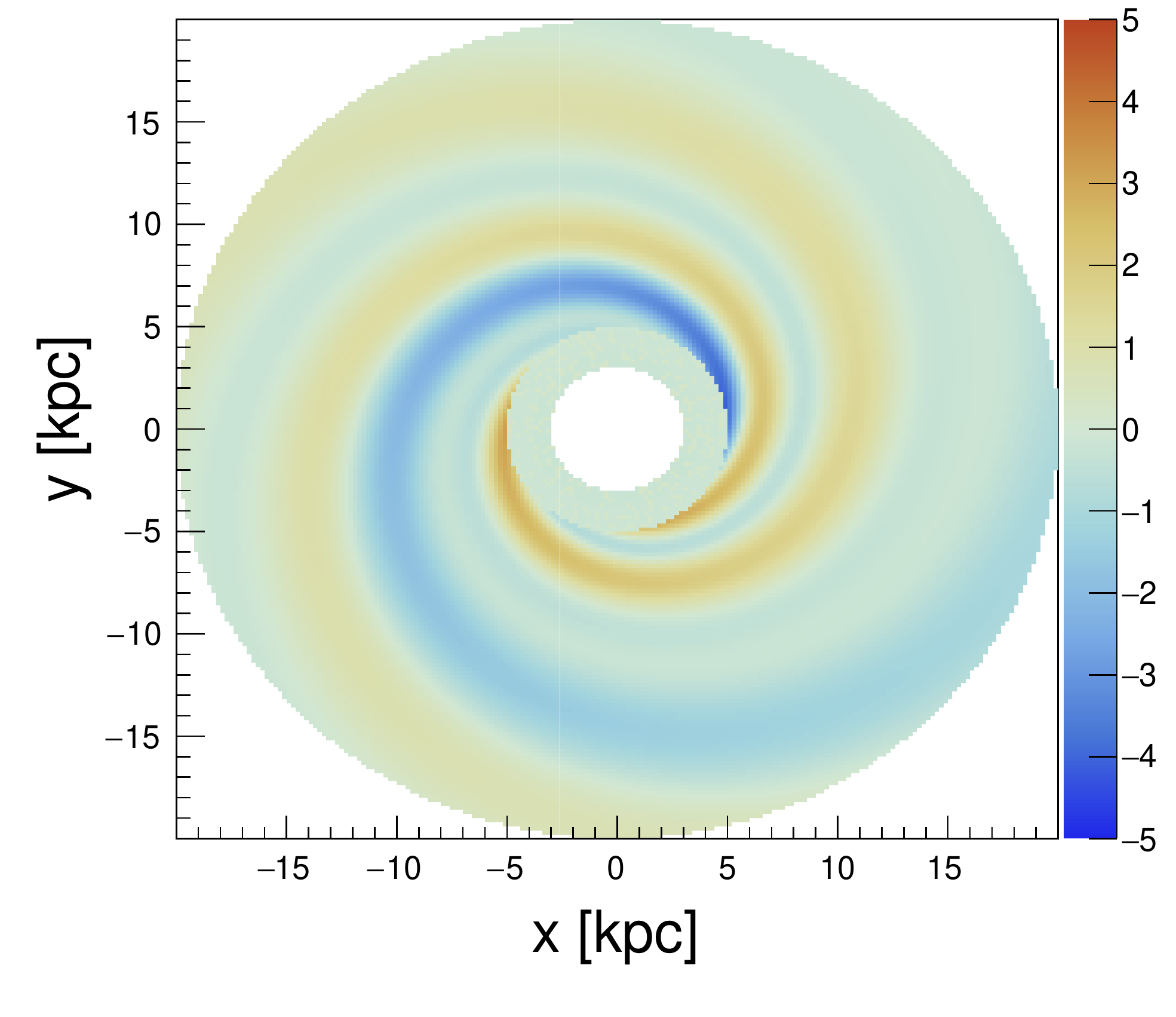}
    \includegraphics[width=0.49\linewidth]{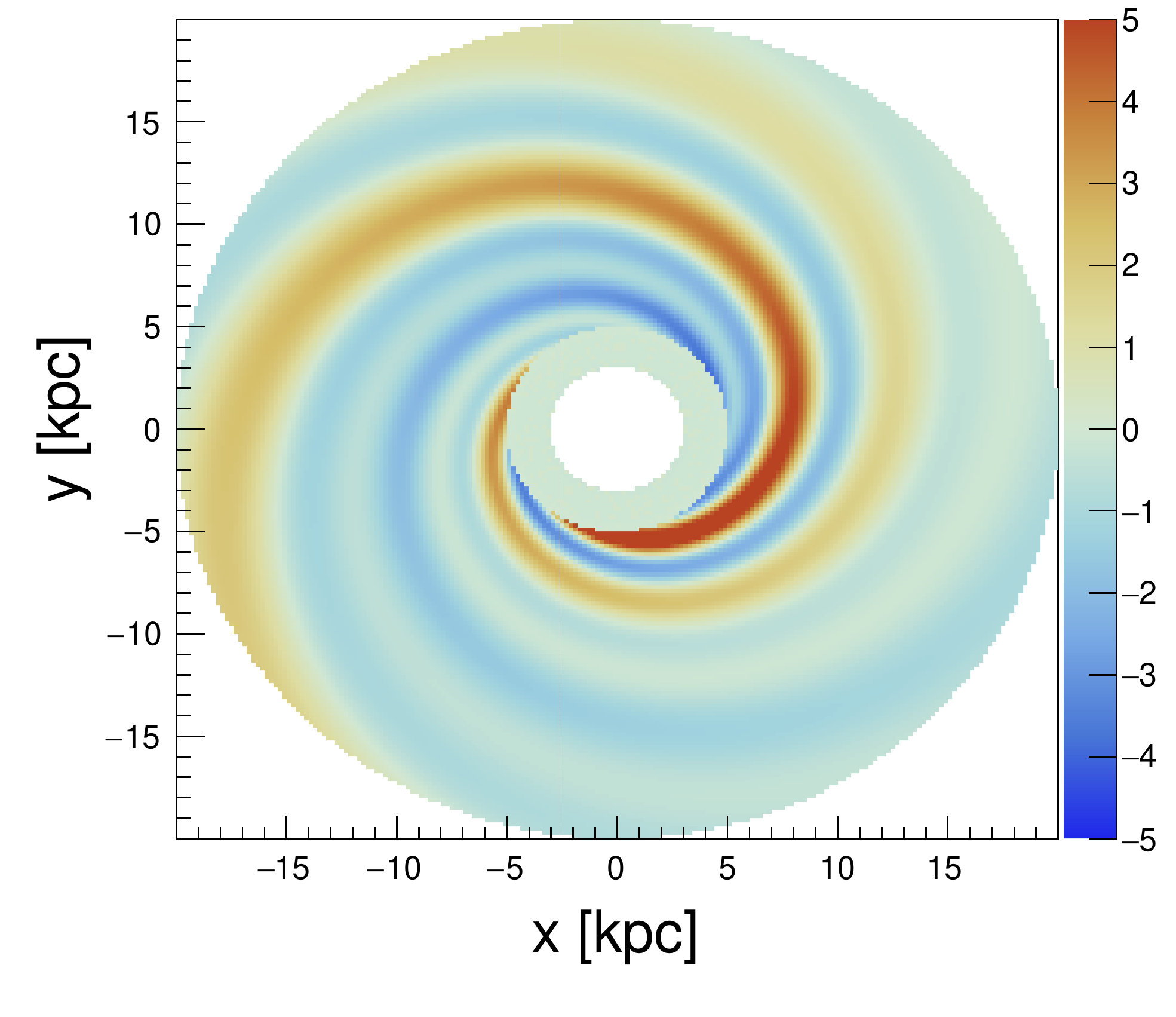}
\caption{Fits of the Disk Field. Upper Left: ``Wedge''-model of Brown+07 with fixed pitch angle of $\alpha=11.5^\circ$. Upper Right: Smooth model with three modes ($\alpha=(13.4\pm0.7)^\circ$). Lower Middle: Smooth model with four modes ($\alpha=(14\pm0.5)^\circ$).}
\label{fig:spiral}
\end{figure}

The spiral model of JF12 follows~\cite{2007ApJ...663..258B} to
describe the disk field as a logarithmic spiral with pitch angle
$\alpha$ ($r = r_0 \exp\left((\phi-\phi_0)\tan\alpha\right)$ and
distinct spiral segments (cf.\ upper left panel in
Fig.~\ref{fig:spiral}). These segments introduce discontinuities of the
magnetic field when passing from one spiral to the next, moreover the arm geometry was not allowed to vary in the JF12 fitting.
To achieve a
smoother transition between arm- and intra-arm regions and facilitate fitting the arm geometry, we instead decompose
the magnetic field strength at reference radius $r_0$ into modes $m_i$
with phase $\phi^*_i$:
\begin{equation}
  B(r_0) = \sum_{i=1}^n B_i\, \cos\left(m_i(\phi_0 - \phi^*_i)\right)
\end{equation}
with which the magnetic field at $(r, \phi)$ is given by
\begin{equation}
  {\bf B}(r, \phi) = \left(\sin \alpha, \cos \alpha, 0\right) \, \frac{r_0}{r}\, B(r_0).
\label{eq:smoothspiral}
\end{equation}
A spiral with three modes (upper right panel of Fig.~\ref{fig:spiral})
improves the fit quality with respect to the baseline model with a
discontinuous spiral by $\Delta\chi^2=166$ (the baseline model has
$\chi^2/\text{ndf} = 7292/6585$).  Four modes yields an additional
improvement of $\Delta\chi^2=18$ (lower panel of
Fig.~\ref{fig:spiral}). Adding more modes does not improve the fit
significantly. The pitch angle of the logarithmic spiral is a free
parameter of the fit and we obtain $\alpha=(13.4\pm0.7)^\circ$ and
$(14\pm0.5)^\circ$ for $n=3$ and $4$ respectively, i.e.\ the pitch
angle of the magnetic field is similar to but not the same as that of other probes of the spiral arms
(see e.g.~\cite{2014ApJ...783..130R}) in accordance of observations of
magnetic fields in nearby galaxies~\cite{2015ApJ...799...35V}. Further
refinements of the JF12 spiral model are underway taking also into
account rotation measures from Galactic pulsars with known distances
(see also~\cite{hanUHECR}).

\subsection{Flaring Disk Field}
Observations of the 
distribution of \HI, CO and stars are compatible with an
exponential increase of the vertical scale height $h$ of the Galactic
disk~\cite{2014ApJ...794...90K}, with increasing distance from the Galactic center:
\begin{equation}
  h(r) = h_0\,e^\frac{r}{r_h}
\end{equation}
with $h_0=0.13$~kpc and $r_h=9.22$~kpc for \HI~\cite{2009ARA&A..47...27K}.
Adopting the same behavior for the vertical scale height of the disk field
induces a vertical component:
\begin{equation}
   B_z(r, \phi, z) = B_r(r, \phi, z) \, \frac{z}{r_h}.
\end{equation}

\begin{figure}[!t]
\centering
\includegraphics[width=0.9\linewidth]{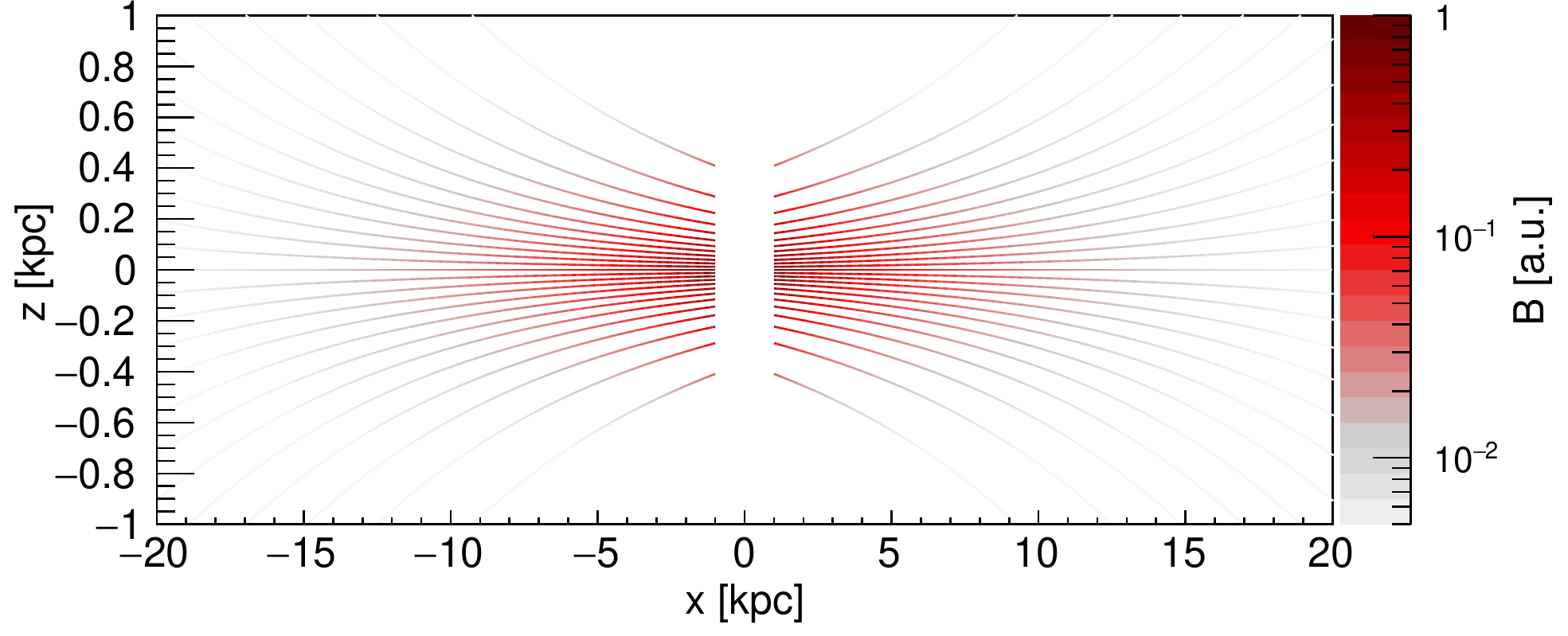}
\caption{Side view of a exponentially flaring radial disk field
  ($h_0=0.13$~kpc and $r_h=9.22$~kpc).}
\label{fig:flare}
\end{figure}

An illustration of a flaring radial disk field ($B_r(r, \phi, z)~1/r$)
is shown in Fig.~\ref{fig:flare}. Applying the flaring to the JF12
model with a freely floating radial scale $r_h$ converges to very
large values of $r_h$, i.e.\ our preliminary conclusion is that the
data does not support a strong flaring of the magnetic disk field, but
further studies are needed for a definite conclusion.

\subsection{Galactic Warp}
It is well known that the Galaxy disk is not perfectly planar but
considerably warped, possibly caused by the gravitational pull of the
Magellanic clouds on the Milky Way. Adding the Galactic warp
from~\cite{2006ApJ...643..881L} to our model, changes the fit quality
by only $\Delta\chi^2=8$, and therefore it can be concluded that the
current data is not very sensitive to the warp. However, the warp
changes the low-latitude magnetic field towards the outer Galaxy and
is therefore important for the backtracking of UHECRs in this
direction.

\begin{figure*}[!t]
  \includegraphics[clip, rviewport = 0 -0.1 1.2 1, width=0.33\linewidth]{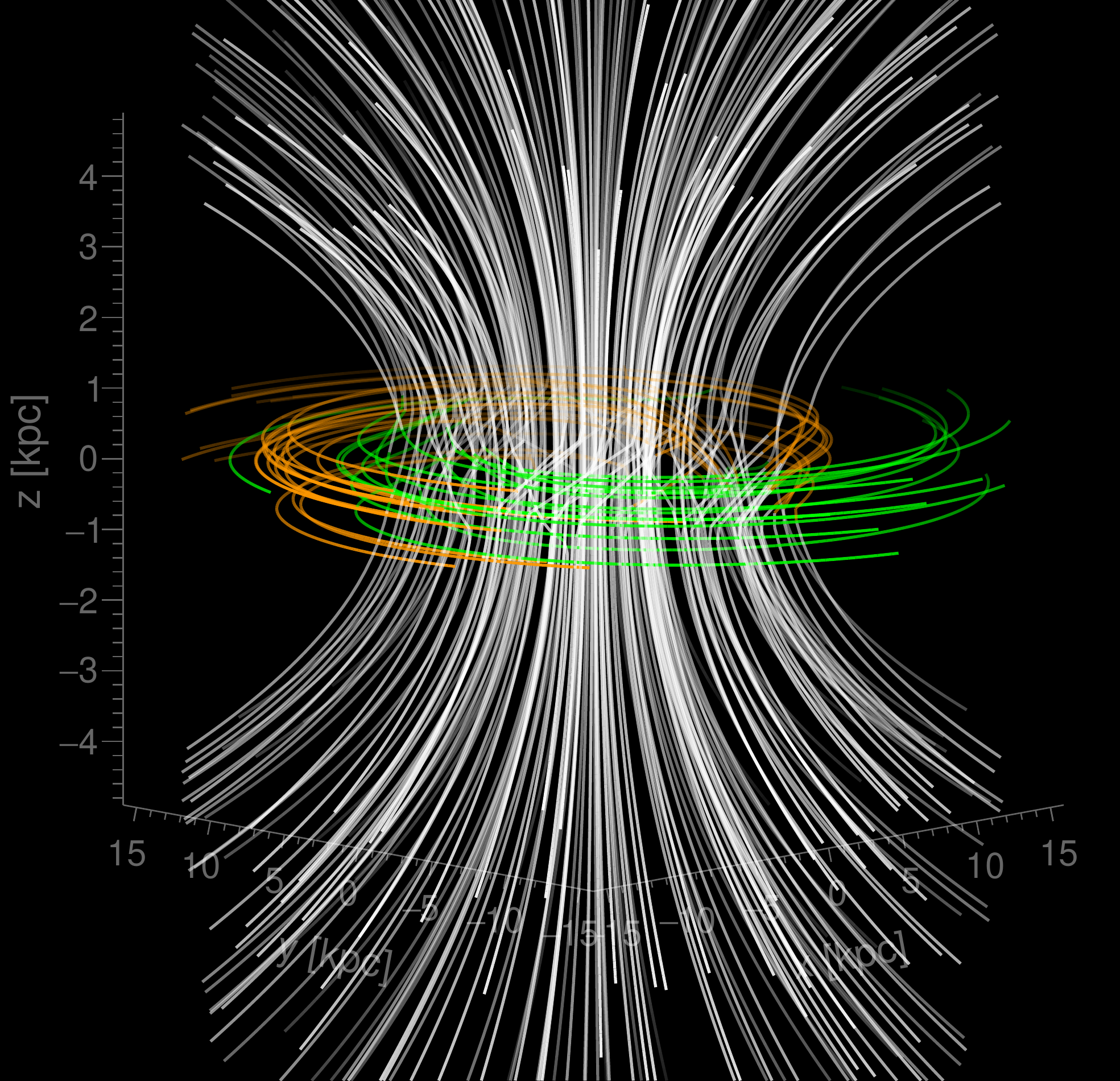}
\includegraphics[clip, rviewport = 0 -0.1 1.2 1, width=0.33\linewidth]{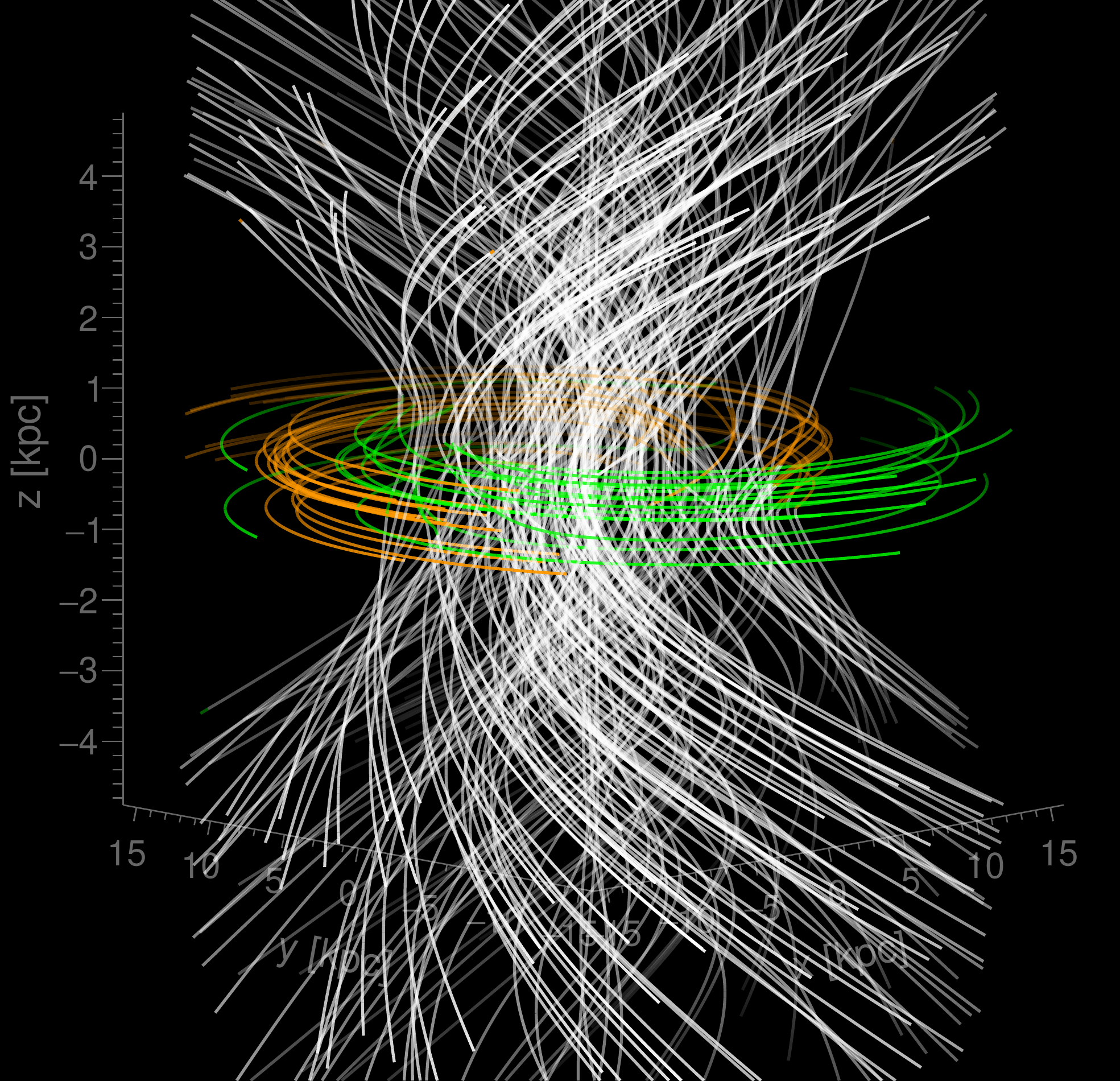} \hspace*{-0.8cm} \includegraphics[clip,rviewport=0 -0.15 1 1,width=0.33\linewidth]{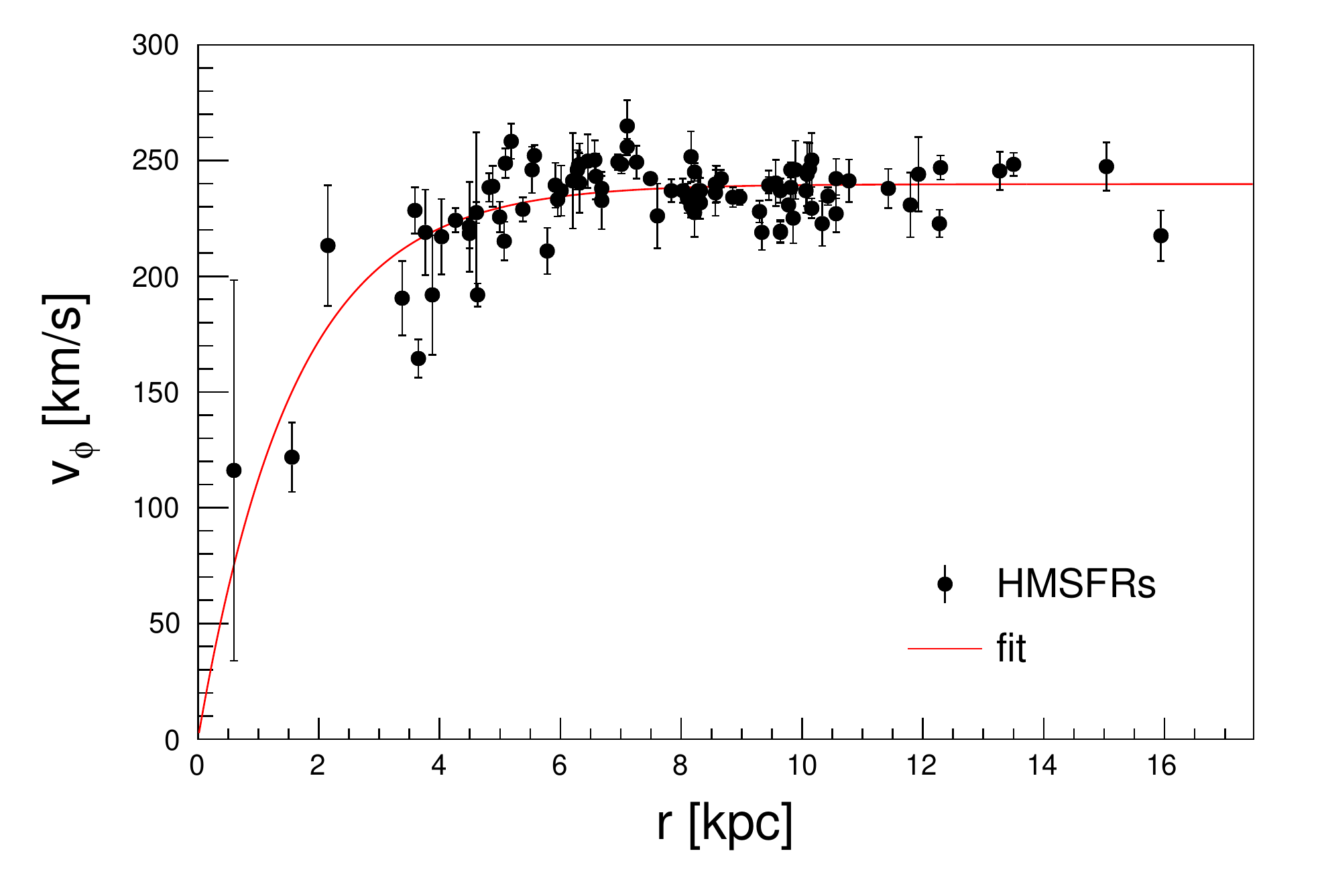} \\
\includegraphics[clip, rviewport=0.5 0 1 1,width=0.33\linewidth]{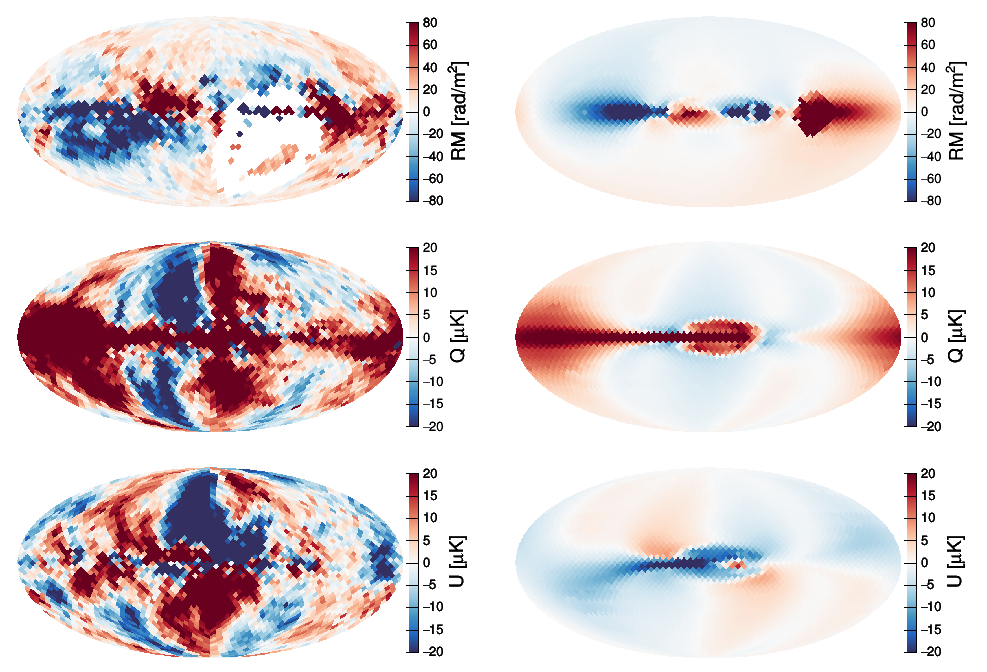}
\includegraphics[clip, rviewport=0.5 0 1 1,width=0.33\linewidth]{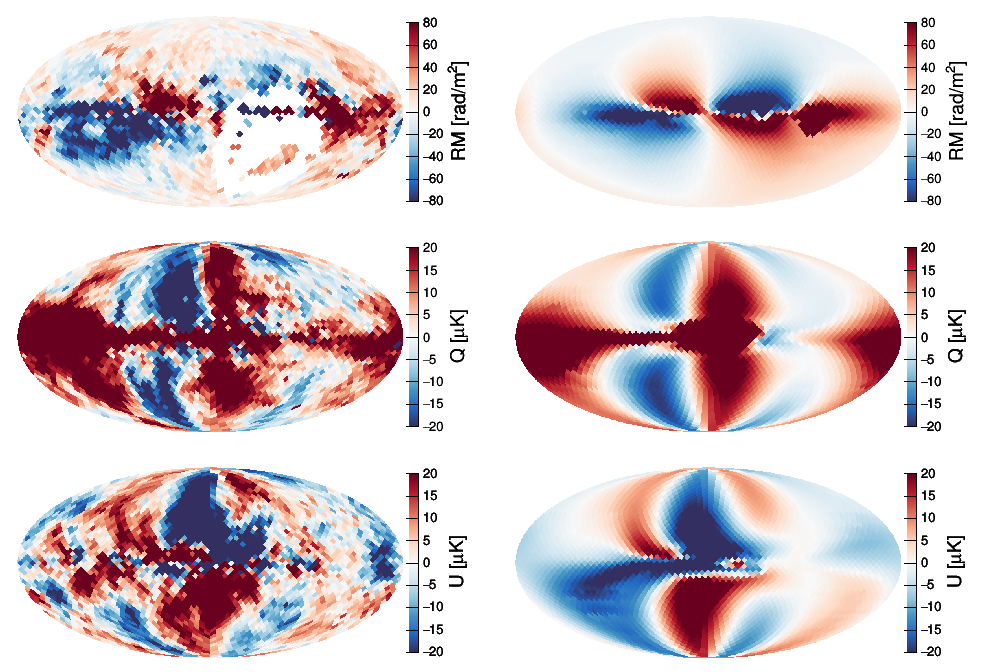}
\includegraphics[clip, rviewport=0 0 0.5 1,width=0.33\linewidth]{pics/uf17_70.png}\\
 \hspace*{1.3cm}(a)\quad$t = 0$~Myr  \hspace*{3.4cm}
(b)\quad$t = 70$~Myr  \hspace*{3.4cm}
(c)\quad data  \hspace*{1cm}
 \caption{Illustration of the GMF model with a twisted poloidal
   field. The best-fit model at $t = 70$~Myr is presented in the
   middle column and the corresponding un-twisted model at $t=0$ is
   shown in the left column. The lower three rows show the simulated
   sky maps for RM, Q and U. A visualization of field lines of the
   models is in the top row. The lines are colored if the toroidal
   field strength is $\geq 99\%$ of the total field strength.  Regions
   with ${\bf{r}} \cdot {\bf{B}}>0$ are shown in orange, otherwise
   green is used.  The observed sky maps are displayed in the lower
   three panels of the right-most column (Q and U data
   from~\cite{2011ApJS..192...15G}, RM data compilation
   from~\cite{2012ApJ...757...14J} (see references therein)).. he
   rotation curve of the Galaxy used for the twisting is shown as a
   red line in the top-right panel together with measured velocities
   of high-mass starforming regions (HMSFRs)
   from~\cite{2014ApJ...783..130R}.}
\label{fig:twist}
\end{figure*}

\subsection{Twisted X-Field}
As noted in~\cite{Farrar:2014hma}, the directions of the toroidal halo
components derived in the JF12 model (in the same direction as the
Galaxy’s rotation in the south and opposite to it the north) are the
directions which would result from differential rotation of the
poloidal component. In the JF12 model the toroidal and poloidal
components were fitted independently, but if indeed the toroidal
component is produced from the poloidal by differential rotation,
there is a more detailed relationship between the radial and vertical
behavior of their field strengths.


Consider an X-field that is dragged along with the rotation of the
Galaxy.  The MHD evolution of a magnetic field
is~\cite{1979cmft.book.....P}
\begin{equation}
  \partial_t {\bf B} = {\bf \nabla} \times ({\bf v} \times {\bf B})
      - \underbrace{{\bf \nabla} \times \eta ({\bf \nabla} \times {\bf B})}_{=0\;{\rm for}\;\sigma\rightarrow\infty}
\end{equation}
where the ``frozen-in condition'' applies for infinite conductivity
(magnetic diffusivity $\eta \rightarrow 0$ when $\sigma \rightarrow \infty$). Under these conditions and for a  purely azimuthal rotation 
\begin{equation}
\partial_t {\bf B} =   {\bf \nabla} \times ({\bf v} \times {\bf B}) =  \begin{pmatrix}
   -\frac{v}{r} \,\partial_\phi B_r \\
    \partial_z (v B_z) + \partial_r (v B_r) \\
    -\frac{v}{r} \, \partial_\phi B_z
\end{pmatrix}
\end{equation}
Thus for a magnetic field that is poloidal and
azimuthally symmetric at $t=0$, $B_\phi$ (only) evolves with time:
\begin{equation}
 {B_\phi(t) = \left(B_z\, \partial_z v + rB_r \,\partial_r \omega\right)\, t},
\label{eq:evolution}
\end{equation}
where we introduced the angular velocity
$  {\bm \omega} = \omega \, {\bf e}_\phi = \frac{v}{r}  \, {\bf e}_\phi$
and used the solenoidality of the poloidal field.

Eq.~(\ref{eq:evolution}) can be applied to evolve any type of poloidal
field. For definiteness, we tested this ansatz by evolving the
smooth poloidal field model of type ``C''
from~\cite{2014A&A...561A.100F}.  For the Galactic rotation curve we
used a fit to the high-mass starforming regions with parallax
measurements from~\cite{2014ApJ...783..130R} (see top right panel of
Fig.~\ref{fig:twist}) and for the vertical velocity gradient we assume
a constant value inspired by simulations~\cite{2014A&A...566A..87J}
and constrain it within two sigma of the value of (22$\pm$6)
(km/s)/kpc as observed close to the Galactic
midplane~\cite{2008ApJ...679.1288L}. The resulting sky maps of RM, Q
and U of the un-twisted model and the evolved model ($t=70$~Myr) are
shown in the left and middle panel of Fig.~\ref{fig:twist}.  In
contrast to the conclusions of~\cite{2003AcASn..44S.151M}, we find a
good description of the overall structure of data from the combined
effect of radial and vertical shear, in particular the anti-symmetric
pattern of the rotation measures and the tilted pattern of Q and U
within the Solar circle.  (The analytics underlying the contrary conclusion of~\cite{2003AcASn..44S.151M} are not apparent to us.)  In a future work we will discuss the magnitude and physical origin of the effective winding time $\approx 70$ Myr.

\subsection{Thermal Electron Model}
A large-scale model of the density of thermal electrons in the Galaxy
($n_e$) is needed to predict the rotation measures for a given
magnetic field configuration. The spatial distribution of $n_e$ can be
estimated using dispersion measures of Galactic pulsars and scattering
measures of Galactic and extragalactic sources. We have tested the impact
of two different models for the thermal electron densities:
NE2001~\cite{2002astro.ph..7156C} with the updated scale height of
the thick disk from~\cite{2008PASA...25..184G} (used in JF12), and
YMW17~\cite{2017ApJ...835...29Y}. The newer YMW17 model benefits from
more available dispersion measures from pulsars with measured
distances, but the more important difference between the two models is
their particular parametric choices for the model components, such as
for instance the thickness and pitch angles of the spiral arms. Both
models give a similar good description of the observed dispersion
measures of pulsars. The main difference between the models
for inferring the  magnetic fields is the different electron density
in the halo. A larger magnetic field strength in the halo is inferred
using YMW17 due to the lower density of electrons at large Galactic height in this model.
The fitted field strengths for the toroidal and poloidal components are
about a factor two larger than if 
derived using NE2001.

In addition to studying the influence of the warm ionized medium on
the GMF modeling, we also tested the effect of adding the large-scale
hot ionized medium from~\cite{2015ApJ...800...14M} to the calculation
of RMs. No significant change of the inferred GMF was found in this
case.

\subsection{Correlation of $n_e$ and $B$}
In the 
baseline fit for the parameters of the GMF  
the magnetic field and thermal electron density 
are taken to be 
uncorrelated. However, as pointed out in~\cite{2003A&A...411...99B}, pressure
equilibrium between the ISM plasma and the magnetic field can lead
to an anti-correlation between the two quantities, leading to an underestimation
of the magnetic field inferred from RMs. 
By contrast, compression leads
to an enhancement of both magnetic field and gas density and results in a positive
correlation of the two quantities,
leading to an overestimation
of the magnetic field inferred from RMs if this is not taken into account.

An approximate relation between the uncorrelated rotation measure
$\RM^{0}$ and the one observed in the presence of correlations was
derived in~\cite{2003A&A...411...99B} and reads as
\begin{equation}
\RM=\RM^{0}
\left(1+\sfrac23\,\kappa\,\frac{\langle\bb^2\rangle}{\BB^2+\langle\bb^2\rangle}\right),
\end{equation}
where $\bb$ and $\BB$ denote the random and coherent magnetic field
strength, respectively, and $\kappa=-1$ for pressure equilibrium and
$\kappa=1$ for compression.

To study the effect of the two extreme cases $\kappa=\pm1$, we
implemented a combined fit of the random and coherent field strength.
As expected, very different coherent magnetic field strengths are
inferred for the two cases of $\kappa=\pm1$.  The total energy of the
coherent magnetic field in the Galaxy is $4\times 10^{55}$~erg for
$\kappa=-1$ and $4\times 10^{54}$~erg for $\kappa=+1$. Under the
standard assumption of no correlation ($\kappa=0$), the coherent
energy is $1\times 10^{55}$~erg. It is worthwhile noting that these
are upper bounds on the effects of a possible $n_e$-$B$ correlation,
because a) the assumed correlation coefficients are at the extreme
values and b) the synchrotron product used in the comparison is the
baseline one from WMAP7 without a spinning dust component, whose
inferred random magnetic field is largest (cf.\ Sec.~\ref{sec:syn}).

\subsection{Cosmic-Ray Electrons}
The density and spectrum of cosmic-ray electrons depends in two ways
on the Galactic magnetic field: Firstly, the GMF determines the
diffusion of the electrons from their sources through the Galaxy and,
secondly, synchrotron losses in the GMF are the main cause of electron
cooling apart from inverse Compton scattering above $\sim 10$~\GeV.  The
JF12 fits used a two-dimensional $n_{\rm cre}$ model from a {\scshape
  GalProp} simulation with a uniform isotropic diffusion coefficient
within a cylindrical volume of 4~kpc height.  We have tested the three
variants of the cosmic-ray electron densities used
in~\cite{2016A&A...596A.103P}, which are updated versions of the
calculations described in~\cite{2010ApJ...722L..58S}
and~\cite{2013MNRAS.436.2127O}. The vertical extent of the diffusion
volume of these simulations is 4~kpc and 10~kpc respectively. The
inferred field strengths of the coherent GMF are relatively
insensitive to these changes, because of the flexibility in the fit to
adjust the relative scale between the RMs and the polarized intensity
by changing the amount of striated fields, i.e.\ aligned anisotropic
random fields that contribute to the polarized intensity, but not to
the rotation measures. Further studies concerning the impact of
cosmic-ray electrons are under investigation, in particular the effect
of using a more detailed three-dimensional source distribution of
relativistic electrons and interstellar radiation fields.

\subsection{Synchrotron Data Products}
\label{sec:syn}
The parameters of the original JF12 model were inferred by using the
7-year WMAP synchrotron maps~\cite{2011ApJS..192...15G}.  We have
investigated the impact on the GMF fit, of using new synchrotron
products provided in the 9-year final WMAP data
release~\cite{2013ApJS..208...20B} and the Planck 2015 data
release~\cite{2016A&A...594A..10P}. These products differ in the
constraints applied to the measured Galactic microwave emission data
to extract the synchrotron component. WMAP7 and the ``base'' model of
WMAP9 fit the data with a sum of synchrotron, free-free and dust
emission. All other models include a spinning dust component to
describe the ``anomalous microwave emission''. Moreover, different
constraints to the spectral index of the synchrotron emission are
applied in different products to extrapolate the prior from the
low-frequency intensity map at 408~MHz~\cite{1982A&AS...47....1H}.

Only small differences in the inferred coherent magnetic field are
found for the various products on the polarized emission.  However, as
already noted in~\cite{2016A&A...596A.103P}, the different treatment
of the anomalous microwave emission in the models used for the
different synchrotron products strongly affects the estimated total
synchrotron intensities, and thus the random magnetic field component.
This difference leads to a reduction of random field strength, by up
to a factor of four in the disk, relative to JF12.  Efforts to better determine the synchrotron component are ongoing.
\begin{figure*}[!t]
    \centering
  \includegraphics[width=0.85\linewidth]{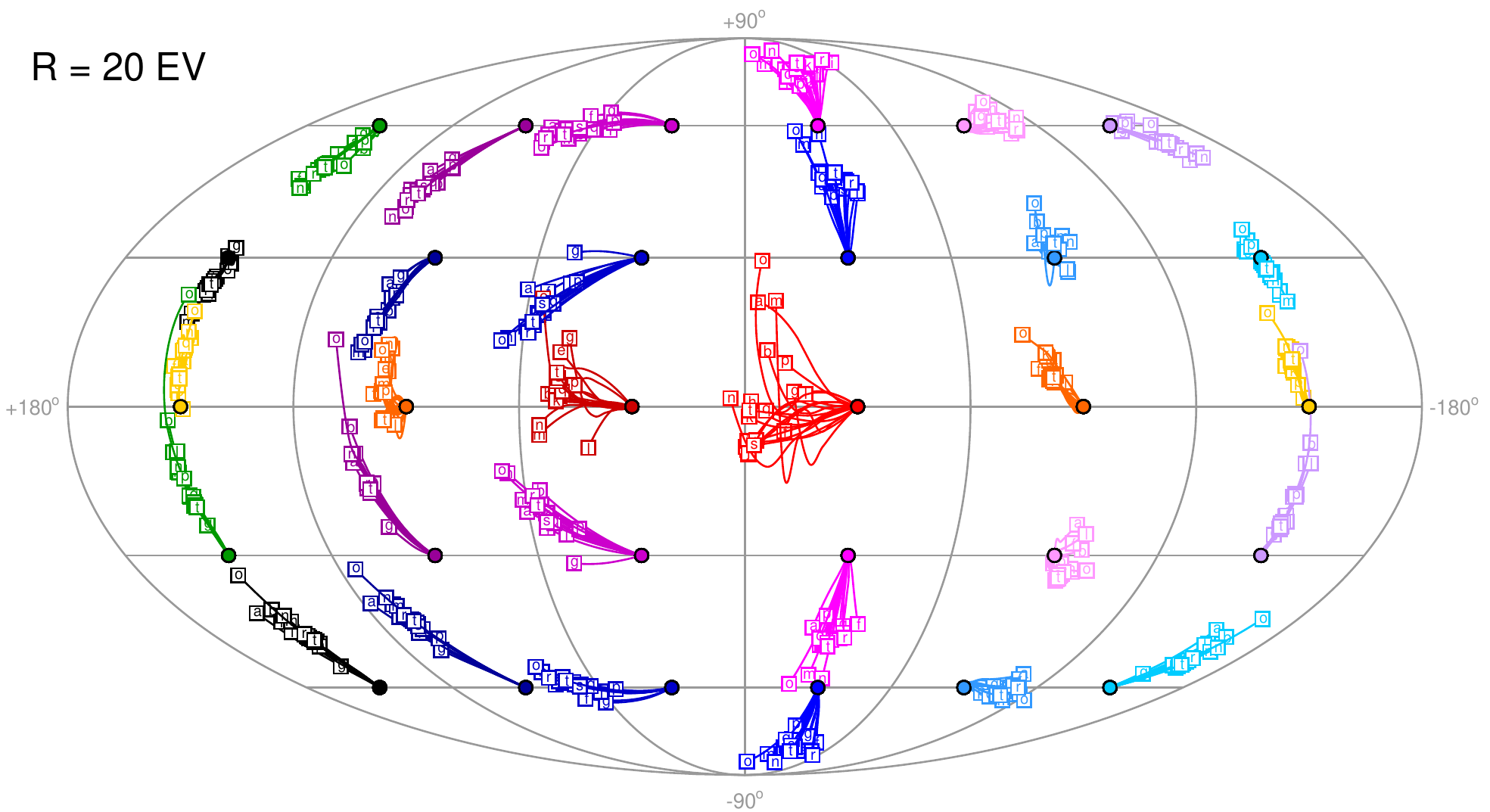}
  \caption{Backtracking of charged particles through the Galaxy
    starting from a regular grid of initial directions (dots).  The
    resulting directions outside of the Galaxy for particles with a
    rigidity of 20~EV are denoted by squares and the lines connecting
    the initial and final positions were constructed by performing
    backtracking at higher rigidities. Each of the letters (a)-(t)
    denotes a different GMF model that describes the synchrotron and
    RM data. }
\label{fig:defl}
\end{figure*}
\begin{figure*}[!ht]
  \centering
  \includegraphics[width=\linewidth]{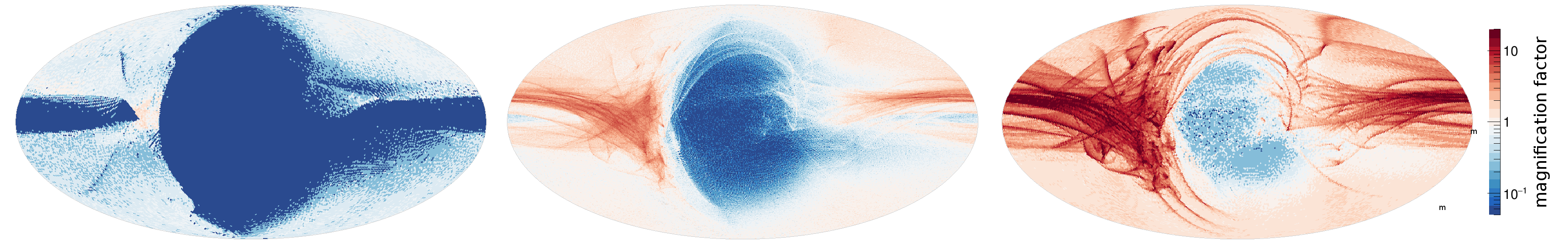}
\caption{Minimum (left), average (middle) and maximum (right)
  magnification factor of the 20 GMF models studied in this work at a
  rigidity of 10~EV.}
\label{fig:ampl}
\end{figure*}

\section{Effect on UHECR Propagation}
The variations and improvements described in the last section lead to
different models of the GMF that all describe the RM, Q, U and I data
well and can thus be regarded as currently-viable descriptions of the
large-scale structure of the GMF.  In total we constructed 20
variations of the JF12 model (see Tab.~1 in~\cite{Unger:2017wyw} for
more details) and studied the impact of these models on the
propagation of ultrahigh-energy cosmic rays in the Galaxy by
backtracking charged particles from Earth to the edge of the Galaxy.
The results are shown in Fig.~\ref{fig:defl} for particle rigidities
$R\geq20$~EV (rigidity = energy/charge). As can be seen, the
deflections predicted by different models are mostly confined within
well-defined regions on the sky. If the variations studied in this
work were bracketing the extreme possibilities for the large-scale
configuration of the GMF, then one could use these results to
construct a correction for the spatially varying average deflection
based on all models and apply it to the arrival directions of UHECRs
together with the uncertainty given by the spread of different model
prediction. However, the variations studied here are most probably not
exhaustive and thus give only a lower limit on the current uncertainty of
the inferred arrival direction of cosmic rays at the edge of the the
Galaxy. Nevertheless, in era of high-statistics data from the Pierre
Auger Observatory and Telescope Array, the time seems ripe to use our
current knowledge of the large-scale structure of the GMF to enhance
studies of correlations between astrophysical sources and the arrival
directions of cosmic rays (see also~\cite{2016APh....85...54E}).

In this context it is also worthwhile noting that not all
extragalactic sources can be observed at Earth, because there exists
no trajectory through the GMF that would allow a charge particle of
certain rigidity to reach Earth from that direction \cite{Farrar:2014hma,faks15,fs17}. This effect can be conveniently
described by the {\itshape magnification factor}~\cite{Harari:2002dy,
  Farrar:2014hma} of the flux arriving from a particular
direction. Sky maps of the magnification factor can be constructed by
performing many backtrackings from Earth on a fine isotropic angular
grid and counting the number of trajectories leading from each
extragalactic arrival direction to Earth \cite{Farrar:2014hma,faks15,fs17}. The obtained magnification
map satisfies Liouville's theorem, since an isotropic flux of
extragalactic cosmic rays will always lead to an isotropic flux at
Earth by construction.

The minimum, average and maximum magnification factors of the 20
coherent GMF models studied in this work are shown in
Fig.~\ref{fig:ampl} for $R=$~10~EV.  As can be seen, the
the GMF optics can cause magnification or demagnification of the
extragalactic flux received at Earth.  Especially for sources behind
and below the Galactic center (as seen from Earth), the 20 model
variations predict unanimously a large demagnification of the flux
arriving at Earth (blue region in the right panel of
Fig.~\ref{fig:ampl}).  This is a general consequence of the poloidal
component of the coherent field \cite{faks15,fs17} which is quite
strong in the central region of the galaxy and persistently present in
all 20 GMF models.

\begin{figure}[!t]
  \centering
  \includegraphics[width=\linewidth]{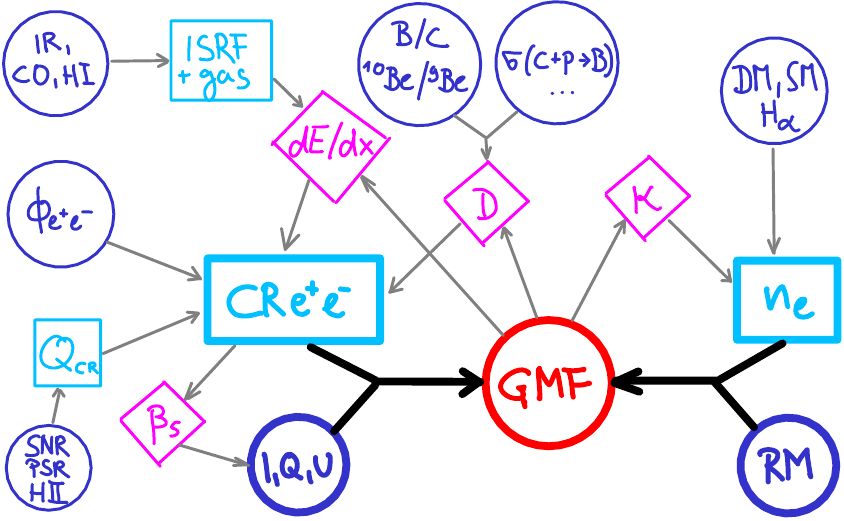}
  \caption{Sketch of the procedure to estimate the GMF from data (blue circles),
    models (azure squares) and auxiliary quantities (magenta diamonds). See text
    for further explanation.}
\label{fig:comp}
\end{figure}

\section{Summary}
In these proceedings we have reported on progress in modeling the
large-scale structure of the coherent and random magnetic field of the
Galaxy. In addition to introducing new parametric models for the
large-scale structure of the coherent GMF, we have studied the
uncertainties of the underlying model assumptions inherent in the
analysis as sketched in Fig.~\ref{fig:comp}.

In the first level of inference, the GMF is estimated from Q,U\&I and
RM data (thick blue circles), using models of the thermal and
cosmic-ray electron density (thick azure boxes).  These models depend
on further data (blue circles: DM, SM and H$_\alpha$ for $n_e$ and
electron+positron fluxes at Earth, $\Phi_{e^+e^-}$, for $n_{\rm
  cre}$). The interpretation of RM relies on an assumption about the
correlation of $n_e$ and $B$, which introduces a feedback loop into
the inference process (magenta $\kappa$ diamond). Another feedback
loop is added by energy loss of cosmic-ray electrons in the Galaxy,
$\text{d}E/\text{d}x$, which is in part due to synchrotron cooling in
the GMF. Further model-dependencies of the energy loss are introduced
by interstellar radiation fields (inverse Compton) and gas
(ionization). In addition, a model of the spacial distribution of
cosmic-ray sources, $Q_{\text{CR}}$, (based on sparse data on SNRs and
their tracers), is needed to calculate three-dimensional density of
cosmic-ray electrons. Even the data itself is not free of model
assumptions, as some of the the estimates of the synchrotron emission
Q,U\&I use the spectral index of synchrotron radiation $\beta_S$,
which in turn depends on the energy spectrum of cosmic-ray
electrons. The diffusion coefficient $D$ and its energy dependence is
a direct effect of the GMF and in principle the diffusion coefficient
could be self-consistently derived for any model of the GMF. However,
usually $D$ is estimated from data on the secondary to primary ratios
of the fluxes of cosmic-ray nuclei (e.g.\ B/C and
$^{10}\text{Be}/^{9}\text{Be}$ given fragmentation cross sections
$\sigma$ in collisions of cosmic-ray nuclei with protons and helium of
the ISM).

The 20 model variations presented here constitute a to-our-knowledge
unique attempt to quantify the uncertainties of large-scale models of
the GMF. Although our study does not necessarily cover the full range
of models compatible with the data, our preliminary conclusion is that
the derived GMF models vary significantly, but that for a typical
application such as the backtracking of ultrahigh-energy cosmic rays,
the data constrains the models enough to encourage further studies of
the possibility of deriving GMF models with a realistic uncertainty.

\section*{Acknowledgments}
{ We would like to thank Tess Jaffe for providing the
  simulations for the cosmic-ray electrons models
  from~\cite{2016A&A...596A.103P}.  MU acknowledges the financial
  support from the EU-funded Marie Curie Outgoing Fellowship, Grant
  PIOF-GA-2013-624803 and the hospitality of CCPP/NYU, where part of
  this research was conducted. The research of GRF is supported in
  part by the U.S.\ National Science Foundation (NSF), Grant
  NSF-1517319.}

\setlength{\bibsep}{0pt plus 0.3ex}
\small
\bibliography{mf}

\end{document}